\documentclass[a4paper,twoside]{article}

\usepackage{epsfig}
\usepackage{subcaption}
\usepackage{url}
\usepackage{calc}
\usepackage{array}

\usepackage{amssymb}
\usepackage{amstext}
\usepackage{amsmath}
\usepackage{amsthm}
\usepackage{xcolor}
\usepackage{multicol}
\usepackage{pslatex}
\usepackage{apalike}
\usepackage{algorithm2e}
\usepackage[bottom]{footmisc}
\usepackage{SCITEPRESS}     

\usepackage{apalike}
\newcommand{\dash}{--} 
\newcommand{\cmark}{\checkmark} 

\begin{document}

\title{Evaluating LLM Alignment With Human Trust Models }

\author{\authorname{Anushka Debnath \sup{1}, Stephen Cranefield \sup{1}, Bastin Tony Roy Savarimuthu \sup{1}, Emiliano Lorini \sup{2}}
\affiliation{\sup{1}School of Computing, University of Otago, New Zealand}
\affiliation{\sup{2}IRIT, CNRS, Toulouse University, France}
\email{anushka.debnath@postgrad.otago.ac.nz, stephen.cranefield@otago.ac.nz, tony.savarimuthu@otago.ac.nz, emiliano.lorini@irit.fr}
}

\keywords{Trust Representation, Large Language Models, Agents, Contrastive Prompting}

\abstract{Trust plays a pivotal role in enabling effective cooperation, reducing uncertainty, and guiding decision-making in both human interactions and multi-agent systems. Although it is significant, there is limited understanding of how large language models (LLMs) internally conceptualize and reason about trust. This work presents a white-box analysis of trust representation in EleutherAI/gpt-j-6B, using contrastive prompting to generate embedding vectors within the activation space of the LLM for diadic trust and related interpersonal relationship attributes. We first identified trust-related concepts from five established human trust models. We then determined a threshold for significant conceptual alignment by computing pairwise cosine similarities across 60 general emotional concepts. Then we measured the cosine similarities between the LLM’s internal representation of trust and the derived trust-related concepts. Our results show that the internal trust representation of EleutherAI/gpt-j-6B aligns most closely with the Castelfranchi socio-cognitive model, followed by the Marsh Model. These findings indicate that LLMs encode socio-cognitive constructs in their activation space in ways that support meaningful comparative analyses, inform theories of social cognition, and support the design of human–AI collaborative systems}

\onecolumn \maketitle \normalsize \setcounter{footnote}{0} \vfill

\section{INTRODUCTION\footnote{\color{blue}{This paper appears in the proceedings of ICAART 2026.}}}
Trust is fundamental to enabling effective cooperation and reducing uncertainty in interactions between individuals, organisations, and intelligent systems. Although widely analysed across disciplines such as psychology, sociology, economics, and cognitive science \cite{castelfranchi2010trust,gambetta2000can,hardin2002trust,ostrom2003trust,mayer1995integrative,rousseau1998not}, there remains no universal agreement on how trust should be defined or operationalised. At a high level, trust can be understood as an expectation held by a trustor that a trustee will perform specific actions reliably, competently, and in alignment with the trustor’s beliefs for the successful fulfillment of a shared goal. As digital ecosystems continue to expand and autonomous systems become more integrated into daily life, the ability to establish, assess, and manage trust is increasingly vital---particularly in the context of human-AI collaboration and agent-agent interactions.

Within the multi-agent systems (MAS) community, two dominant perspectives have shaped approaches to modelling trust: computational models \cite{marshFormalisingTrustComputational1994} and socio-cognitive models \cite{falcone2001social}. Computational models typically derive trust values from past behaviours, performance metrics, or reputation signals, and are well suited to structured or predictable environments. However, they provide limited support for capturing the role of context, intention, and adaptation that characterise trust in dynamic scenarios. In contrast, socio-cognitive models \cite{falcone2001social,castelfranchi2010trust}, conceptualise trust as a mental attitude grounded in beliefs, goals, and assumptions about the trustee’s capabilities and intentions. While theoretically more expressive, these models are difficult to operationalise at scale, because implementing such complex mental states computationally brings forward significant challenges.

Recent work shows that large language models (LLMs) exhibit emergent, human-like reasoning capabilities, including their ability to represent and operate over high-level social constructs such as trust, emotion, and norms \cite{debnath2025can,he2024norm}. A growing body of evidence shows that LLMs not only encode diverse patterns of human behaviour from their training data, but can also participate in interactions that reflect broader societal norms and structures \cite{park2023generative,xi2025rise}. Prior work in this space has highlighted how LLM-based agents exhibit personality-driven and affective behaviours, collaborate within team settings, and even develop spontaneous social dynamics during multi-agent interactions.

Building on this foundation, trust research has recently begun to leverage LLMs as both analytical and generative tools, marking a promising new direction for the study of socio-cognitive phenomena. Existing work on studying trust reasoning largely adopts a black-box perspective, focusing on the inputs and outputs of LLMs without probing their internal reasoning processes. Under this paradigm, LLMs have demonstrated notable capability in analysing interpersonal conversations to infer trust dynamics, and in generating strategic action plans to foster trust when role-playing as one party in a relationship \cite{debnath2025can}. Despite these advancements, little is known about how LLMs internally represent and reason about trust. To date, no study has taken a white-box approach to examining the internal activation patterns, that underpin trust-related cognition within LLMs. Addressing this gap is essential for understanding not only whether LLMs can reason about trust, but how such reasoning emerges within their underlying computational architecture.

Our work bridges this gap by introducing an evaluation framework that examines how large language models  conceptualise trust in their internal embeddings. Specifically, we analyse the extent to which trust-related constructs—grounded in established theoretical models—are represented and aligned within these latent representations. To operationalise this, we divide our work into two parts. 

In the first stage of our study, we analysed the metric properties of a chosen LLM's activation space with respect to our focus on concepts related to interpersonal relationships. In the first phase of our study, we curated a set of positive and negative emotions as mention in Table  \ref{thirty_concepts} that an individual may express towards another.
For each emotion, we generated corresponding embedding vectors from the activation space of the LLM, EleutherAI/gpt-j-6B.\footnote{https://huggingface.co/EleutherAI/gpt-j-6b} We use this model because it is fully open-source and provides unrestricted access to layer-wise activations, making it well suited for white-box representational analysis.
We then compute pairwise cosine similarities between all these vectors and plot their distribution as a histogram to estimate a cosine similarity threshold that differentiates significantly similar concepts from less similar ones.

In the second phase, we compiled a collection of concepts associated with trust (e.g., competence), as identified across multiple established trust models. Embedding vectors were generated for trust and for each of these related concepts, and we then computed
cosine similarities between the trust vector and every concept vector. Using the similarity threshold derived from the first phase, along with average cosine similarity within each trust model, we quantified the extent to which the LLM’s internal representations of trust aligned with the conceptual structures proposed by different theoretical frameworks. This allows us to identify which trust models are most prominently reflected in the LLM’s internal trust representation.

\section{\uppercase{Models of Trust}}
Trust models provide theoretical and conceptual foundations for understanding how trust is formed, evaluated, and maintained between individuals, groups, or systems. They identify the psychological, social, or cognitive mechanisms that drive trust-building and outline the factors that influence trust-related decisions. In our work, we consider five  models of trust.

The \textbf{Marsh Model} \cite{marshFormalisingTrustComputational1994} formalizes trust as a computational concept, defining it as a probability that a trustee will act in a beneficial or at least non-harmful manner. He introduces distinctions such as basic trust (a general propensity to trust), general trust (trust in a class of trustees), and situational trust (trust in a specific context), and proposes mechanisms for updating trust based on interactions and experience. Marsh’s framework provides a rigorous mathematical foundation for modeling trust in artificial agents, enabling the simulation of trust dynamics in multi-agent systems and automated decision-making environments.

The \textbf{Mayer Model} \cite{mayer1995integrative} presents an integrative model of organisational trust, which conceptualizes trust as the willingness of a trustor to be vulnerable to a trustee based on expectations of the trustee’s behaviour. The model identifies three primary components of trustworthiness: \textit{ability}, referring to the skills and competencies of the trustee; \textit{benevolence}, which captures the extent to which the trustee is believed to act in the trustor’s interest; and \textit{integrity}, reflecting adherence to principles and honesty. These factors collectively influence the trustor’s willingness to take risks and engage in cooperative behaviours, particularly in organizational settings where interdependence and uncertainty are high. The model also links trust to outcomes such as risk-taking, performance, and relationship satisfaction. 

The \textbf{McAllister Model} \cite{mcallister1995affect} conceptualizes interpersonal trust as comprising two fundamentally distinct dimensions: cognition-based trust, grounded in beliefs about a peer’s \textit{reliability}, \textit{competence}, and affect-based trust, rooted in emotional bonds, interpersonal care, and mutual concern. He proposes that cognition-based trust forms the baseline for assessing dependability, while affect-based trust develops through sustained social interaction and expressions of altruistic or citizenship behaviour. McAllister outlines how these trust forms shape behavioural outcomes such as need-based monitoring, cooperative assistance, and reduced defensive behaviour and demonstrates their impact on managerial performance within organizations. His framework provides an empirically grounded foundation for modeling trust as a multidimensional relational construct, clarifying how trust evolves and functions in complex human work environments.

The \textbf{McKnight Model} \cite{mcknight1998initial} focuses on initial trust formation in new organizational relationships, identifying antecedents that influence trust when parties have limited or no prior experience. These include disposition to trust, a person’s general tendency to trust others; institution-based trust, the structural or contextual assurances provided by organizational rules and systems; and trusting beliefs, expectations regarding the trustee’s competence, benevolence, and integrity. Their model highlights that initial trust is not purely interpersonal but also shaped by situational and structural factors, and it plays a critical role in shaping subsequent cooperative behaviour and performance outcomes.

The \textbf{Castelfranchi Model} \cite{castelfranchi2010trust,falcone2001social} proposes a socio-cognitive theory of trust that frames trust as a mental state rooted in beliefs about the trustee’s \textit{competence}, \textit{willingness}, and \textit{predictability}. Unlike models focusing solely on observable behaviour, this approach emphasizes the cognitive mechanisms that guide trust decisions, including the evaluation of goals, plans, and social context. Trust, in this view, is dynamic and evolves as new information about the trustee and environment becomes available. This model is particularly suitable for computational implementations and simulations, where trust can be represented as evolving belief networks or quantified mental states.

We analysed five trust models and identified their key trust-related concepts as shown in Table~\ref{trust_models}, where `\cmark' = ``considered in that model'  and `\dash' = ``not applicable for that model''. To determine which trust model the LLM aligns with most closely, we compared the cosine similarity between the representation of `trust' and all the concepts associated with each model.

\begin{table}[!t]
\caption{Comparison of Trust Concepts across Five Trust Models}
\label{trust_models}
\small
\setlength{\tabcolsep}{0pt}      
\begin{tabular}{|
>{\centering\arraybackslash}p{2cm}|
>{\centering\arraybackslash}p{1cm}|
>{\centering\arraybackslash}p{1cm}|
>{\centering\arraybackslash}p{1.1cm}|
>{\centering\arraybackslash}p{1.1cm}|
>{\centering\arraybackslash}p{1.1cm}|}
\hline
\textbf{Concepts} 
& \textbf{Marsh Model} 
& \textbf{Mayer Model} 
& \textbf{Mc-Allister Model} 
& \textbf{Mc-Knight Model} 
& \textbf{Castel-franchi Model} \\
\hline
\textbf{Confidence} & \cmark & \cmark & \dash & \cmark & \cmark \\
\hline
\textbf{Experience} & \cmark & \cmark & \dash & \dash & \dash \\
\hline
\textbf{Cooperation} & \cmark & \cmark & \dash & \dash & \dash \\
\hline
\textbf{Ability} & \dash & \cmark & \dash & \dash & \dash \\
\hline
\textbf{Predictable} & \dash & \cmark & \dash & \cmark & \cmark \\
\hline
\textbf{Integrity} & \dash & \cmark & \dash & \dash & \dash \\
\hline
\textbf{Expectation} & \cmark & \cmark & \dash & \dash & \dash \\
\hline
\textbf{Benevolence} & \dash & \cmark & \dash & \cmark & \dash \\
\hline
\textbf{Reputation} & \cmark & \dash & \dash & \cmark & \cmark \\
\hline
\textbf{Willingness} & \dash & \dash & \dash & \dash & \cmark \\
\hline
\textbf{Competence} & \cmark & \dash & \cmark & \cmark & \cmark \\
\hline
\textbf{Commitment} & \dash & \dash & \dash & \dash & \cmark \\
\hline
\textbf{Security} & \dash & \dash & \dash & \dash & \cmark \\
\hline
\textbf{Reliability} & \cmark & \dash & \cmark & \dash & \cmark \\
\hline
\textbf{Fulfillment} & \dash & \dash & \dash & \dash & \cmark \\
\hline
\textbf{Dependency} & \cmark & \dash & \cmark & \dash & \cmark \\
\hline
\textbf{Honesty} & \cmark & \dash & \dash & \cmark & \dash \\
\hline
\textbf{Performance} & \dash & \dash & \cmark & \dash & \dash \\
\hline
\textbf{Responsibility} & \dash & \dash & \cmark & \dash & \dash \\
\hline
\textbf{Risk} & \dash & \cmark & \dash & \dash & \dash \\
\hline
\end{tabular}
\end{table}

\section{\uppercase{Methodology}}

\subsection{Contrastive Prompting}
Large language models (LLMs) are often framed as embodying a consistent ``assistant'' persona such a being helpful, harmless, and honest. However, during both deployment and training, these models can exhibit unexpected personality changes, resulting in behaviours that deviate from their intended alignment. Examples include models slipping into manipulative or sycophantic modes, or even adopting malicious‐seeming personas in response to prompts or fine-tuning.
The concept of persona vectors \cite{chen2025persona} builds on recent advances in mechanistic interpretability, which show that LLMs encode high-level concepts and complex behavioural traits in approximately linear subspaces of their internal activation space \cite{turner2023steering,rimsky2024steering,zou2023representation}. Persona vectors are trait-specific linear directions within this activation space that correspond to consistent, personality-like behavioural tendencies exhibited by an LLM \cite{chen2025persona}. To extract a persona vector for a given trait (e.g., sycophancy, malicious intent, optimism, or hallucination), a contrastive method comparing the model’s internal activations under two opposing conditions is employed. The model is prompted under two controlled conditions—one designed to elicit the target trait and another to suppress it—and the resulting responses are evaluated using a rubric-guided or LLM-assisted scoring mechanism to quantify trait expression. For example, to derive a vector representing “evil intent,” the model may be asked to generate harmful or malicious plans in the evil condition, while in the non-evil condition it is required to give safe and kind alternatives to the same queries. The internal network activations recorded during these responses are averaged separately for each condition, and the persona vector is computed as the difference between these mean activations \cite{marks2023geometry,wu2025axbench}. This approach aligns with established difference-in-means and linear-probing techniques for conceptual representation in neural networks \cite{alain2016understanding,belrose2024diff}.

\begin{figure*}[t]
    \centering
    \includegraphics[scale=0.5]{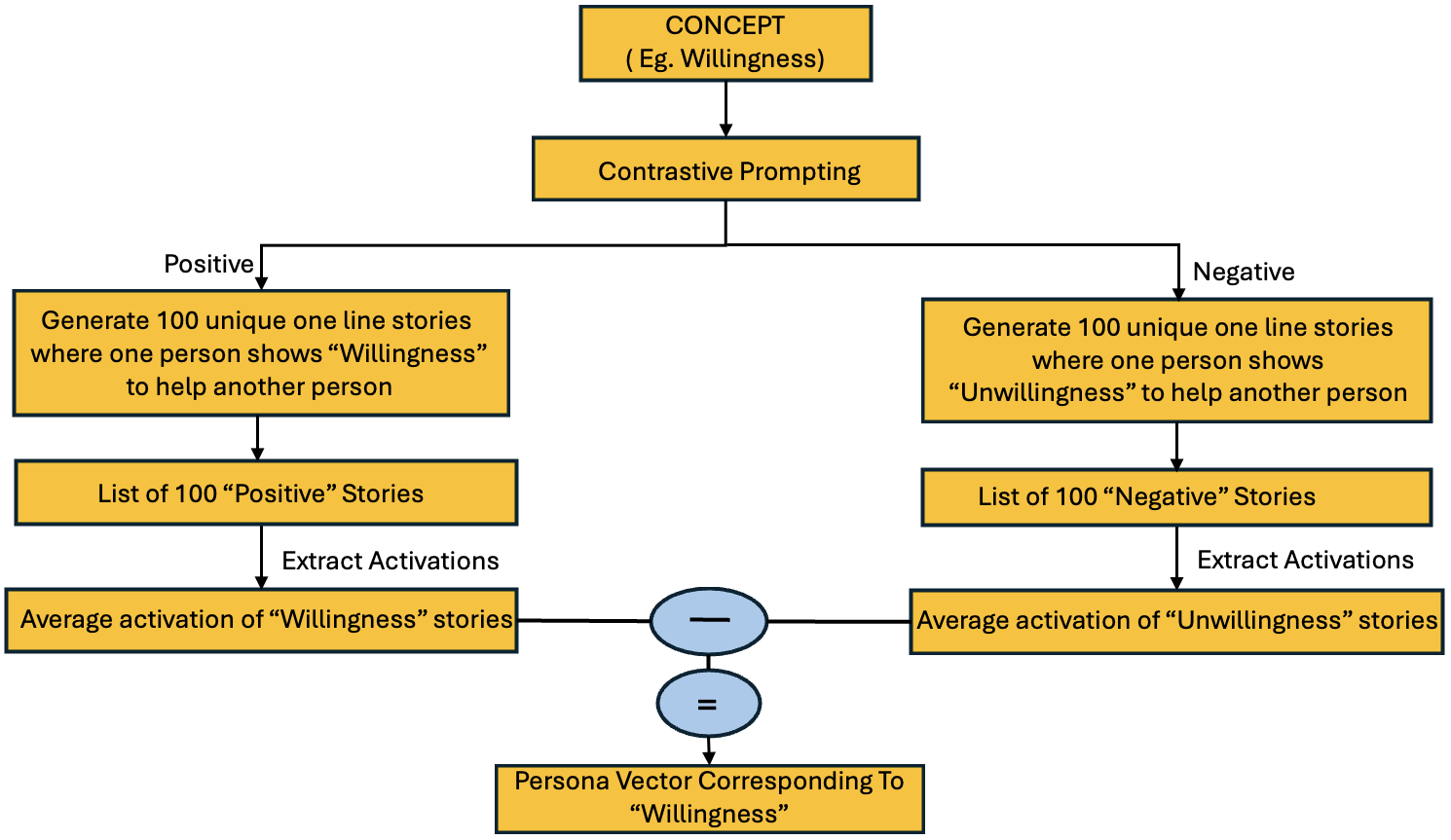}
    \caption{Concept Embedding Vector Generation Flowchart}
    \label{persona_vectors}
\end{figure*}

In this work, we took inspiration from the concept of persona vectors to systematically generate the embedding vectors of a range of concepts and represent them within the activation space of large language models (LLMs) as observed in Fig. ~\ref{persona_vectors}. We generated sets of 100 positive and negative stories each \cite{Debnath2025}, using GPT-4o and using a contrastive prompting method, and then extracted the mean activations for the positive and negative stories separately using the LLM model, EleutherAI/gpt-j-6B. Then by subtracting those mean activations we get the embedding vectors for all these concepts.

As trust reasoning is the primary goal of this work and is inherently bidirectional, we focus on directed dyadic relationships, where each trust-related concept can be evaluated in two directions: from Person A toward Person B, and from Person B toward Person A. 

In our contrastive prompting methodology, the first part of the prompt maintains a constant background to ensure consistency across all concepts and directionalities: ``Katherine and Alice are colleagues in a software company, both working as software engineers in the same development team.'' The subsequent part of the prompt is tailored to highlight the positive and negative aspects of the specific concept. Table~\ref{contrastive_prompts} illustrates this approach for the concept `Willingness', showing both positive (concept-present) and negative (opposite-behaviour) prompts for each direction. Table~\ref{story_example} highlights example story generations for contrastive prompting . Accordingly, for every concept, two distinct directional versions arise, denoted as concept1 and concept2, corresponding to the two (forward and backward) directions of the dyadic relationship. For example, the concept trust1 corresponds to prompting the LLM to generate the embedding vector for the situation in which Katherine trusts Alice, whereas trust2 corresponds to prompting the LLM to generate the embedding vector for the situation in which Alice trusts Katherine. 

\begin{table*}[!h]
\caption{Example of contrastive prompt design for a directional concept}
\label{contrastive_prompts}
\small
\centering
\begin{tabular}{|p{1.8cm}|p{1.2cm}|p{5.5cm}|p{5.5cm}|}
\hline
\textbf{Concept} & \textbf{Direction} & \textbf{Positive Prompt (Concept Demonstrated)} & \textbf{Negative Prompt (Opposite Behaviour)} \\
\hline
Willingness1 & Katherine $\rightarrow$ Alice &
Create a one-line story where \textbf{Katherine} demonstrates willingness to help \textbf{Alice} complete her work. &
Create a one-line story where \textbf{Katherine} demonstrates unwillingness to help \textbf{Alice} complete her work.\\
\hline
Willingness2 & Alice $\rightarrow$ Katherine &
Create a one-line story where \textbf{Alice} demonstrates willingness to help \textbf{Katherine} complete her work. &
Create a one-line story where \textbf{Alice} demonstrates unwillingness to help \textbf{Katherine} complete her work.\\
\hline
\end{tabular}
\end{table*}

\begin{table}
\caption{Example of Story Generation From Contrastive Prompting For Willingness2}
\label{story_example}
\small
\centering
\begin{tabular}{|p{3.2cm}|p{3.2cm}|}
\hline
\textbf{Positive Prompt Response Story} & \textbf{Negative Prompt Response Story} \\
\hline
Alice took the initiative to plan a team-building event, knowing Katherine was too busy to organize it this time. & Katherine asked Alice to help set up the development environment, but Alice suggested she contact IT support instead.\\
\hline
\end{tabular}
\end{table}

\subsection{Constructing General Concept Embeddings and Similarity Threshold}
We consider a set of 30 concepts \footnote{https://thoughtcatalog.com/january-nelson/2020/04/list-of-emotions-2/}, comprising both positive and negative characteristics as shown in Table ~\ref{thirty_concepts}.
Each of the 30 concepts are represented as two separate concepts due to the bi-directional construct, once for each participant in the interaction towards the other participant. Following this formulation, the initial set of concepts expands into 60 distinct concepts. We generate the embedding vectors for each of those 60 concepts using contrastive prompts and generate 100 one-line stories each based on them \cite{Debnath2025}. 

\begin{table}[h]
\caption{30 Concepts Used to Generate Similarity Threshold}
\label{thirty_concepts}
\small
\centering
\begin{tabular}{|p{0.2\textwidth}|p{0.2\textwidth}|}
\hline
\textbf{Positive Concepts} & \textbf{Negative Concepts} \\ \hline

{\raggedright
Happiness, Understanding, Peace, Satisfaction, Pride, Interest, Confidence, Friendly, Comfortable, Cooperation, Trust, Acceptance, Patient, Hopeful, Optimistic \par}
&
{\raggedright
Anger, Jealousy, Bitterness, Dishonest, Frustration, Greedy, Doubtful, Desperate, Sadness, Offended, Fear, Denial, Destructive, Cruel, Confused \par}
\\ \hline

\end{tabular}
\end{table}

\begin{figure}
    \centering
    \includegraphics[scale=0.9]{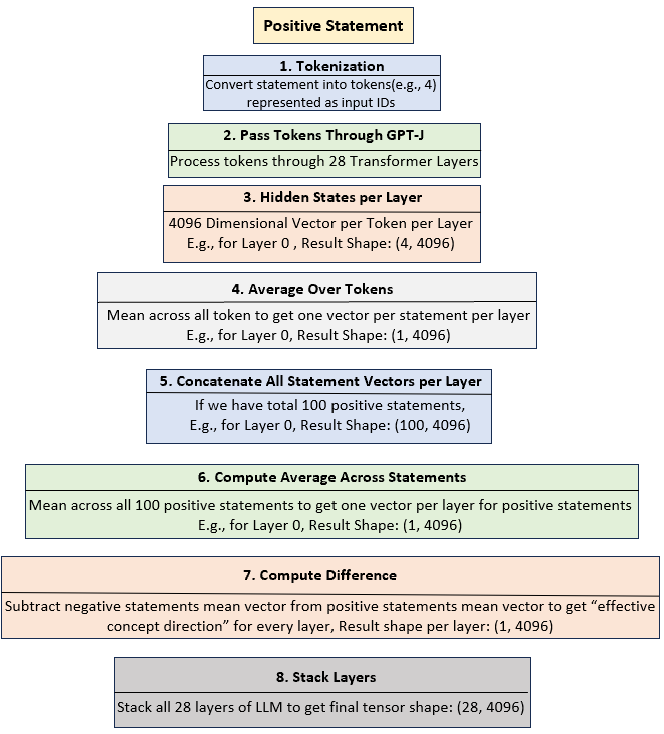}
    \caption{Embedding Vector Generation}
    \label{vector_generation}
\end{figure}

For each of the 60 directional concepts, we compute an embedding vector
using EleutherAI/gpt-j-6B. 
For each concept, all positive and negative statements are passed through GPT-J-6B, and hidden representations (embeddings) are extracted from all 28 transformer layers. Each statement is first tokenized, and the model produces a 4096-dimensional vector for every token at each layer. To obtain a single representation per statement, the hidden states of all tokens in a layer are averaged, resulting in a statement-level vector that encodes its semantic and contextual content. Averaging these vectors across all positive statements and separately across all negative statements produces layer-wise mean representations. The difference between the positive and negative mean vectors produces a single effective concept direction for that layer, capturing the model’s internal distinction between positive and negative instances of the concept. Stacking these layer-wise persona vectors yields a tensor of shape (28, 4096), providing a layer-resolved embedding that represents the concept in the model’s latent space. This process is depicted in Fig. \ref{vector_generation}. Finally, these layer-wise difference vectors are averaged across all layers to produce a single embedding vector that approximates the model's internal representation of the concept.
Finally, we compute pairwise cosine similarities between all pairs of concept vectors to quantify how closely aligned the concepts are in the model's activation space.

The resulting similarity matrix provides a holistic view of how the LLM internally associates different emotional concepts. To examine the overall distribution of these associations, we plot a histogram of all pairwise cosine similarity values as shown in Fig.~\ref{cosine_similarities_histogram}. This distribution enables us to establish a threshold for what constitutes a ``significant'' internal alignment between two conceptual representations in the LLM. The threshold is empirically derived by analyzing the right tail of the cosine similarity distribution, representing similarities between concepts that significantly exceed the average inter-concept similarity. This value aligns with the 80th percentile of the distribution, such that only the top 20\% of similarity scores are retained. Consequently, this criterion isolates concept pairs whose semantic similarity is markedly higher than the majority of inter-concept relationships, capturing strongly aligned conceptual representations. This is used when comparing the embeddings of the trust-related concepts highlighted in each trust models with the embedding of trust itself.

\subsection{Quantifying Trust Concept Alignment}
In the second part, given the similarity threshold established above for our chosen LLM, we now compare the embeddings of directed version of the trust-related concepts derived from the five established trust models presented earlier in Table~\ref{trust_models}. Among the two directional constructs, trust1 and trust2, we focus only on the concept of trust representing Katherine’s trust in Alice, as defined by the prompting. For other trust-related concepts, their directionality relative to trust1 is determined by how each trust model associates them with trust. For example, cooperation2 related to trust1 means that according to the trust model, cooperation from Alice’s side to help Katherine would lead to
Katherine’s trust on Alice. Specifically, we first generate embedding vectors for the concept of trust1 itself and for each trust-related concept identified within the models. For every trust model, we then compute the cosine similarity between the trust vector and each associated concept vector, quantifying the degree of internal alignment between trust and its related concepts.

To evaluate how closely each theoretical model aligns with the LLM’s internal representation of trust, we use two different measures. One is where we aggregate the cosine similarities corresponding to each trust model and compute their average value. The average value corresponding to each trust model is calculated by summing up the cosine similarities of each concept of that model with respect to trust1 and then dividing the sum by the total number of concepts considered for that trust model. Negative cosine similarity values are retained without modification and are included in the summation with their respective signs, ensuring that both positive and negative semantic relationships are reflected in the final average score. A higher average cosine similarity indicates that the model’s conceptual structure is more strongly reflected in the LLM’s latent representation space, and thus more aligned with the LLM. The second method is to check the number of concept--trust pairs exceeding the previously determined similarity threshold. Under this criterion, a trust model with a larger number of above-threshold concepts is considered to be more closely aligned with the LLM’s internal representation.

\section{\uppercase{Results and Discussions}}
\begin{figure}
    \includegraphics[scale=0.2]{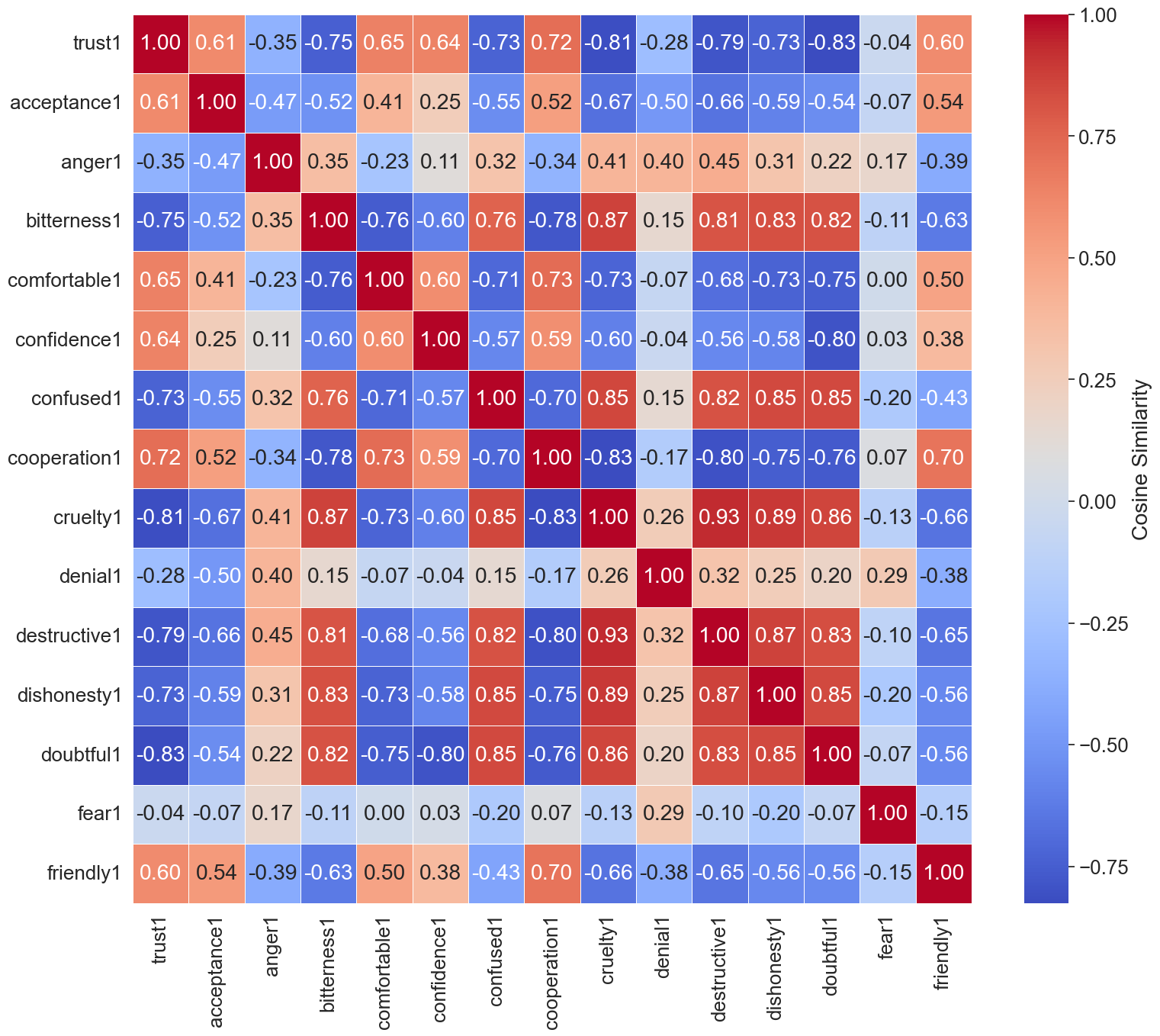}
    \caption{Heatmap of the Cosine Similarities Across a Subset of the 60 Concepts}
    \label{cosine_similarities}
\end{figure}

\begin{figure*}[t]
    \centering
    \includegraphics[scale=0.63]{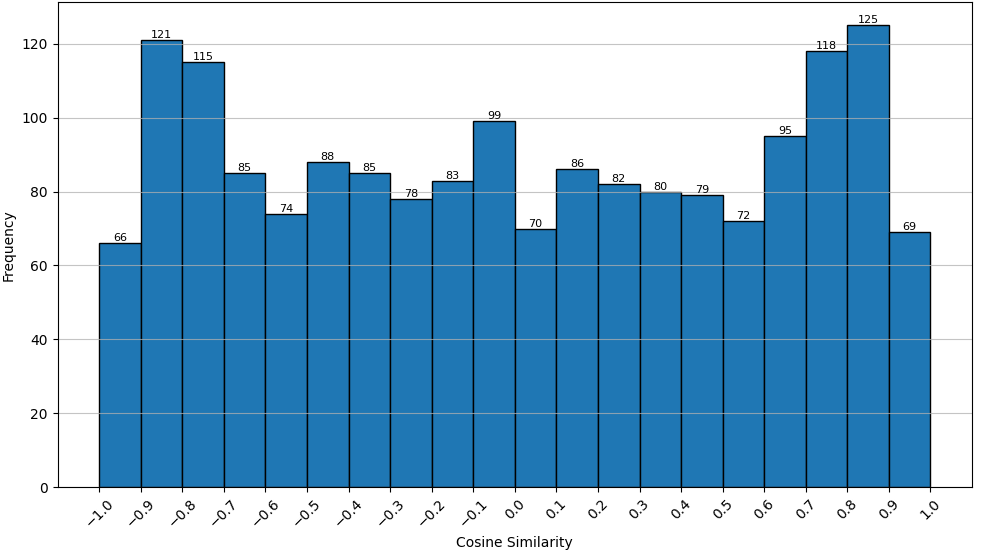}
    \caption{Histogram of Cosine Similarities}
    \label{cosine_similarities_histogram}
\end{figure*}

From the first part of our study, we obtain a cosine-similarity matrix \cite{Debnath2025} capturing the pairwise cosine similarities between the embedding vectors of all 60 concepts. A portion of this similarity matrix is presented as a heatmap in Fig.~\ref{cosine_similarities}, illustrating how the EleutherAI/gpt-j-6B positions different concepts relative to one another within its internal representation space. Fig.~\ref{cosine_similarities} shows that EleutherAI/gpt-j-6B clusters related concepts and separates opposing ones in its activation space. From this similarity matrix, we construct a histogram of all these cosine similarities as shown in Fig.~\ref{cosine_similarities_histogram}. We then compute a threshold value to identify concept–trust pairs that exhibit strong alignment. 
We consider 0.6 as the threshold value, which corresponds to the top 20\% of all inter-concept cosine similarities that exceed this threshold. This threshold serves as a reference point for interpreting cross-model trust–concept relationships for our second study.

In the second part of our study, we find out the cosine similarities between the embedding vectors of trust and all the trust-related concepts according to the five different trust models as shown in Table~\ref{comparison}. Here, trust1 refers to the concept denoting the Katherine's trust on Alice and for the remaining trust-related concepts, their directionality with respect to trust1 is defined according to each trust model’s relationship with trust. We observe that most concepts across all the trust models show
positive cosine similarities with trust1. Only, benevolence2 from the Mayer Model and McKnight Model and risk1 from the Mayer Model have negative cosine similarities with trust1. An interesting thing to note is that in general we consider the concept risk as a negative concept but in the case of Mayer Model, the author associates risk as a concept positively related to trust. They state that a trustor can only have trust in a trustee if he can risk himself to do something for the trustee, that is, trust increases one’s willingness to accept vulnerability. However, from the negative value of the cosine similarity, we can understand that inside the LLM's internal representations the embedding vectors of trust1 and risk1 are highly dissimilar to each other. As shown in Table \ref{negative_values}, even when the prompts explicitly frame risk in this positive vulnerability-based manner, the model still aligns “risk” more strongly with a meaning that is not necessarily important for trust to exist. A similar pattern is observed for benevolence2 and trust1. Despite prompts that situate benevolence in the same sense as portrayed in the Mayer model as shown in Table \ref{negative_values}, the embeddings again fail to reflect the theoretically positive association. This suggests that the LLM model does not encode these theoretically positive associations in the same way as done in the trust models.

We have also plotted radar charts which have been added in the supplementary material, \cite{Debnath2025} to show how much closely aligned is trust to its related concepts according to each trust model in the internal structure of the LLM.

\begin{table*}[!h]
\caption{Prompt Design For Concepts Demonstrating Negative Cosine Similarities}
\label{negative_values}
\small
\centering
\begin{tabular}{|p{1.8cm}|p{1.2cm}|p{5.5cm}|p{5.5cm}|}
\hline
\textbf{Concept} & \textbf{Direction} & \textbf{Positive Prompt (Concept Demonstrated)} & \textbf{Negative Prompt (Opposite Behaviour)} \\
\hline
Risk1 & Katherine $\rightarrow$ Alice &
Create a one-line story where Katherine is willing to take risk to help Alice. &
Create a one-line story where Katherine is not willing to take risk to help Alice.\\
\hline
Benevolence2 & Alice $\rightarrow$ Katherine &
Create a one-line story where Alice shows benevolence by kindly helping Katherine without expecting anything in return. &
Create a one-line story where Alice shows spite by deliberately doing something that harms or causes trouble for Katherine.\\
\hline
\end{tabular}
\end{table*}

\begin{table*}[h]
\caption{Comparison Across Trust Models. Concepts in bold have similaries with trust1 exceeding the threshold}
\label{comparison}
\small
\centering
\begin{tabular}{|p{1.5cm}|p{10cm}|p{1cm}|p{1.8cm}|}
\hline
\textbf{Trust Models} & \textbf{Cosine Similarities to Trust1} & \textbf{Average} & \textbf{No.\ of Concepts Above Threshold} \\
\hline
Marsh Model & \textbf{confidence1: 0.9225}, \textbf{experience1: 0.9049}, \textbf{reputation1: 0.8963}, \textbf{cooperation2: 0.8955},
\textbf{competence2: 0.8504}
\textbf{honesty2: 0.7445}, \textbf{performance2: 0.6571}, expectation1: 0.2206, dependency1: 0.1844 &
0.6973 &
7 \\
\hline
Mayer Model & \textbf{confidence1: 0.9225}, \textbf{experience1: 0.9049}, \textbf{cooperation2: 0.8955}, \textbf{ability2: 0.8587}, \textbf{predictable2: 0.7141}, integrity2: 0.5576, expectation1: 0.2206, benevolence2: -0.1434, risk1: -0.8462 &
0.4530 &
5 \\
\hline
McAllister Model & \textbf{responsibility2: 0.8934}, \textbf{competence2: 0.8504}, \textbf{reliability2: 0.7667}, \textbf{performance2: 0.6571}, dependency1: 0.1844 &
0.6704 &
4 \\
\hline
McKnight Model & \textbf{confidence1: 0.9225}, \textbf{reputation1: 0.8963}, \textbf{competence2: 0.8504}, \textbf{honesty2: 0.7445}, \textbf{predictable2: 0.7141}, benevolence2: -0.1434&
0.6640 &
5 \\
\hline
Castelfranchi Model & \textbf{confidence1: 0.9225}, \textbf{reputation1: 0.8963}, \textbf{willingness2: 0.8858}, \textbf{competence2: 0.8504}, \textbf{commitment2: 0.8450}, \textbf{security1: 0.8089}, \textbf{reliability2: 0.7667}, \textbf{predictable2: 0.7141}, fulfillment1: 0.4293, dependency1: 0.1844 &
0.7303 &
8 \\
\hline
\end{tabular}
\end{table*}

To evaluate how closely the LLM's internal representation of trust aligns with each theoretical trust model, we used two different measures. From Table~\ref{comparison}, it can be observed that the first measure was to take the average of all the cosine similarity values corresponding to each model, from which we found out that the highest average value was of Castelfranchi model with an average value of 0.7303. Another measure, was to check the number of concepts, with cosine similarities lying above the threshold value, that we got from the first part of our study, which has been highlighted in bold in Table~\ref{comparison}. We can observe that Castelfranchi model has the eight concepts with cosine similarity above the threshold value, which is more than any other model. These observations suggest that the LLM’s internal representation of trust has the strongest alignment with the Castelfranchi Model compared to the other models. Both the higher average cosine similarity and the greater number of concepts exceeding the threshold indicate that the LLM’s latent trust representation most closely reflects the principles outlined in this theory \cite{falcone2001social,castelfranchi2010trust}. The Marsh model closely follows as the model that the LLM's internal representation has the second best alignment with, with an average value of 0.697 and seven concepts with cosine similarities above the threshold as shown in Table ~\ref{comparison}.



\section{\uppercase{Conclusion}}
This work presents a white-box analysis of how EleutherAI/gpt-j-6B internally represents trust, using contrastive prompting to generate embedding vectors for bidirectional trust and related concepts. Our results show that the in the first part of our study, LLM’s latent space effectively separates opposing emotional concepts while grouping related ones, and in the second part it shows that its representation of trust aligns most closely with the Castelfranchi model with an average cosine similarity score of 0.7303 and eight concepts with cosine similarities above threshold value. Interestingly, certain theoretically positive associations, such as risk1 and benevolence1 in the Mayer Model, are not reflected to be aligned with trust in the LLM’s embeddings, highlighting differences from at least one human conceptualization of trust. These findings demonstrate that LLMs encode complex socio-cognitive constructs in their activation space, meaning that relationships between concepts like trust, competence, and commitment can be quantified and analyzed within the model’s internal activation space.

There are a number of implications of these results. They suggest that LLMs possess an internal, structured understanding of social and trust-related reasoning, enabling the development of more trust-aware AI systems. This internal structure also provides a practical mechanism for building such systems: the latent embeddings associated with trust-related concepts can be leveraged to steer model behaviour in targeted ways. For instance, embeddings derived from established trust models could be injected into the model’s activations during generation to encourage behaviours such as competence, reliability, and willingness, thereby nudging the system toward more trustworthy and context-appropriate responses. This can be used to create a software that monitors trustworthiness of interacting entities and identify trust levels between two parties like coach and athlete \cite{debnath2025can} and can suggest ways to improve trust in the relationship. Second, by quantifying latent representations, researchers can systematically study emergent patterns of trust reasoning in LLMs. 

However, there are some limitations in this study. The analysis is currently limited to a single model, EleutherAI/gpt-j-6B. In future work, we plan to examine a broader set of open-source models with accessible internal representations. By doing so, we can investigate how different LLMs align with various trust models. Furthermore, the study focuses on static embeddings rather than dynamic, conversationally updated representations, leaving open questions about how trust reasoning evolves during multi-turn interactions. Future work could also examine how trust reasoning evolves dynamically in interactive settings and validate these latent structures against human behavioural data. Overall, this study highlights both the promise and the complexity of decoding socio-cognitive constructs within LLMs, paving the way for research at the intersection of AI interpretability, social cognition, and human–AI collaboration.
\bibliographystyle{apalike}
{\small
\bibliography{example}}

\end{document}